\documentstyle[aps,prl,twocolumn,psfig]{revtex}
\begin{document}

\draft

\preprint{Not yet submitted}

\wideabs{
\title{Periodic Orbit Theory for Rydberg Atoms in External Fields}

\author{P.A.\ Dando, T.S.\ Monteiro and  S.M.\ Owen}
\address{Department of Physics and Astronomy, University College London,
Gower Street, London WC1E 6BT, U.K.}
\date{Submitted: 21 July 1997.  Resubmission: 25 November 1997}
\maketitle

\begin{abstract} 
Although hydrogen in external fields is a paradigm
for the application of periodic orbits
and the Gutzwiller trace formula to a real system, 
the trace formula has never been applied successfully to other atomic
species. We show that spectral fluctuations of general
Rydberg atoms are given  with remarkable precision
by the addition of diffractive terms. 
Previously unknown features in atomic spectra are exposed:
there are new modulations
that are neither periodic orbits nor combinations of
periodic orbits; `core-shadowing' generally 
decreases primitive periodic orbit amplitudes
but can also lead to {\em increases\/}.
\end{abstract}

}

Periodic orbit theory, in the form of the Gutzwiller trace formula 
(GTF)~\cite{G90}, provides the most powerful framework for the
semiclassical quantization of chaotic systems. 
It is more than a decade since it was first shown
that the GTF provides a quantitative description of the oscillations in 
the density of states of highly excited hydrogen atoms in magnetic 
fields~\cite{W87}.
However, the trace formula has never been applied successfully to 
any other species of singly excited (Rydberg) atom. This is
clear from comparisons between accurate quantal spectra 
since the spectral amplitudes for non-hydrogenic atoms 
differ substantially from those of hydrogen~\cite{Jans93}. 

Much effort has been expended in developing
closed orbit theory~\cite{Delos,GD92,B89,A89},
the semiclassical theory which describes 
photoabsorption by atoms in external static fields.  
By matching semiclassical waves to Coulomb waves near the nucleus,
the photoabsorption strength
is obtained as a sum of contributions from only those orbits that
close at the nucleus.  In contrast, 
{\em all\/} periodic orbits contribute to the GTF (the eigenvalue
spectrum).  It was shown
that, provided core-scattered waves are included consistently 
in the matching procedure, closed orbit theory can
be applied to general atoms in fields~\cite{Dando95,Dando96,Shaw96}. 
But, for non-hydrogenic atoms one finds
additional modulations of $O(\sqrt{\hbar})$ 
or higher relative to the
hydrogenic contributions---the `combination-recurrences'---that are 
due to sums of closed orbits. Contributions from 
the harmonics of closed orbits are reduced in amplitude through
`core-shadowing'~\cite{Dando95,Dando96,Shaw96} but those associated
with the first traversals of the primitive orbits are unaffected.

However, closed orbit theory does not account, even qualitatively, 
for the observed differences between
atomic species for the eigenvalue spectrum, {\em i.e.\/} the density of 
states, $\rho(E) =-{\rm Im}\, {\rm Tr}\, G(E)/\pi$. 
In the density of states---quantitatively well-described by periodic orbit 
theory for hydrogen---amplitudes of
primitive periodic orbits can vary substantially between atoms, in 
contrast with closed orbit theory. Also, the
modulations associated with combinations of periodic orbits appear at 
different orders in $\hbar$ relative to closed orbit theory.

Rydberg atoms and molecules in the field-free case are described
by Quantum Defect Theory (QDT), one of the most widely used theories
in atomic physics. In QDT, the effects of a
multi-electron core are described by a set of phase-shifts, 
or `quantum defects', $\delta_{l}$,
in each partial wave, $l$. In the limit when the quantum defects vary 
smoothly with, $l$, they can be related to the {\em classical\/}
precession angle of the Kepler ellipse by $\Theta = 2\pi d\delta_{l}/ dl$. 
For many atoms only the lowest partial waves have 
non-zero quantum defects.  For example,  for even parity lithium, 
$\delta_{0} \simeq 0.4\pi$ and $\delta_{l \geq 2}\simeq 0$ 
while for helium $\delta_{0}\simeq 0.3\pi$ and $\delta_{l \geq 2} \simeq 0$.
In this case, variation of $\delta_{l}$ with $l$ is 
clearly not smooth.

In this letter, we present an approach that, for the first time, 
combines the Gutzwiller trace formula with Quantum Defect Theory
and hence sheds new insight on the classical
interpretation of quantum defects. Our approach yields simple analytical 
expressions for the differences in amplitudes for general non-hydrogenic 
atoms. We compare the new semiclassical results with full quantal calculations 
and, for well isolated orbits, find them to be extremely accurate, for 
example to within about $1\%$ at $\hbar \simeq 0.01$.

A surprising finding, predicted by theory and confirmed by 
quantal results, is that, although the amplitudes of the primitive orbits 
are mostly reduced, as one would expect from the idea of `core-shadowing',
in the non-hydrogenic case they can also
{\em increase\/}.  This is shown below to be due to a de-phasing between 
diffractive (core) and geometric (Coulomb) contributions. Combinations
of periodic orbits appear with order at least $\hbar$. Most significantly, 
new modulations appear that are not combinations of real periodic orbits 
but are rather pure diffractive orbits. They pass through the core and are  
made periodic by the diffraction. We emphasize that all these effects are 
accurately described by the diffractive periodic orbit theory.

The periodic orbit theory of diffraction was developed recently for 
Hamiltonians with discontinuities~\cite{Vatt,Primack,Whelan}.  
For our purposes, a good example of a diffractive system is the
cardioid billiard, which has a single sharp vertex. In this case, periodic 
orbits are decomposed into two kinds:  those that do not intersect the 
vertex (geometric orbits) and those that do (diffractive orbits). 
The  density of states has been shown to be obtained as a sum:
\begin{equation}
\rho(E) = -\frac{1}{\pi} {\rm Im} \,{\rm Tr}\, G_g(E) - 
\frac{1}{\pi} {\rm Im}\, {\rm Tr}\, G_D(E).
\end{equation}
The first (geometric) term yields the ordinary GTF. The trace,
over the second (diffractive) contribution has been shown to 
be~\cite{Vatt,Whelan}
\begin{equation}
{\rm Tr}\,G_D(E)=   \sum_p \frac{T_p}{i\hbar}\prod_n d(n) G(q_n,q_{n+1};E),
\label{diffrac}
\end{equation}
where $T_p$ is the total sum of periods taken over the paths between 
the vertices and $d(n)$ is the diffraction constant which depends on 
the type of diffraction.  Equation~(\ref{diffrac}) 
encapsulates the important result 
that the trace integral taken between the $n$\/th and $n+1$\/th vertices
is proportional to the Green's function between those points. 

We apply diffractive periodic orbit theory to our atomic systems
 by treating the non-hydrogenic core as a diffractive source.
The crucial step is to obtain an expression for the diffractive constant, 
$d$ in terms of quantum defects.  To this end, we consider an incoming 
Coulomb wave, 
$\psi^{(-)}_{\rm Coul}$, which approaches the atomic core from infinity 
at an angle, $\theta_f$, to the $z$-axis.  On reaching the core, 
$\psi^{(-)}_{\rm Coul}$ produces a scattered wave, $\psi_{\rm scatt}$, 
which feeds outgoing semiclassical waves along periodic orbits;
$\psi_{\rm scatt}$ can be decomposed  into 
an outgoing Coulomb wave together with a core-scattered wave~\cite{GD92}:
$\psi_{\rm scatt}(r, \theta) = \psi^{(+)}_{\rm Coul}(r, \theta) 
+ \psi^{\theta_f}_{\rm core}(r, \theta)$.
The Coulomb scattered wave is strongly back-focussed along 
$\theta\simeq\theta_f$ and can be written in closed form~\cite{GD92}.
Our first approximation consists of equating $\psi^{(+)}_{\rm Coul}$
with the source for geometric paths ({\em i.e.\/} the usual GTF). 
The core-scattered wave $\psi^{\theta_f}_{\rm core}$, arising from the 
incoming wave at angle $\theta_f$,  is equated with the source of 
diffractive  semiclassical waves.  At a radius, $r_0\simeq 50\,$Bohr, 
we express $\psi^{\theta_f}_{\rm core}$ in a partial-wave 
expansion which, for $m=0$ is~\cite{GD92}
\begin{equation}
\psi_{\rm core}^{\theta_f} = 
\left( \frac{2\pi^2}{r^3} \right )^{\frac14}
\sum_{l=0}^{\infty}  
Y_{l0}^{*}(\theta_f,0) Y_{l0}(\theta,0)
e^{-i 3\pi/4 }
( e^{2i\delta_l} - 1 )
\label{theory.7}
\end{equation}
where $\delta_l$ are the quantum defects.
Finally, we take $d$ to be the fractional amplitude scattered by the 
core:
\begin{equation}
d(\theta_i,\theta_f)= \psi_{\rm core}^{\theta_f}(r_0,\theta_i)/
\psi^{(-)}_{\rm Coul}(r_0,\theta_f). 
\end{equation}
 
All calculations and comparisons with  fully quantal spectra presented 
here have been carried out for $s$-wave scattering (appropriate for
atoms such as lithium or helium which are used frequently in experiments of
atoms in fields). So, below
$\delta \equiv \delta_{0}$ and,   
in this case, $\psi^{\theta_f}_{\rm core}$ is isotropic.  
However, generalization to odd parity or atoms with multiple 
quantum defects is straightforward. 

We consider the specific example of Rydberg atoms in a static magnetic
field of strength, $\gamma$ (atomic units).
The quantum spectra are calculated at
constant scaled energy $\epsilon=E \gamma^{-2/3}$, that is
for {\em fixed\/} classical dynamics. 
Quantum mechanically,  we calculate a set eigenvalues,
$\gamma_i^{-2/3}$, corresponding to different effective 
$\hbar$~\cite{DD94}.  Below, $\hbar$ denotes $\gamma^{1/3}$.

For Rydberg atoms in a magnetic field, the best studied periodic orbits 
are the straight line orbit perpendicular to the field, $R_1$, and 
the `balloon' orbit, $V^1_1$.  The well-known 
Garton-Tomkins orbit~\cite{GT69}, $R_1$, is responsible for the
quasi-Landau oscillations observed in $m=1$  
atomic spectra near the ionization limit at energy spacing
$\sim 1.5\hbar \omega$;  these were  the first observed `footprints'
of periodic orbits in a real physical system. The balloon orbit
dominates $m=0$, odd-$l$ spectra with oscillations of 
spacing $\sim 0.64\hbar \omega$.  The
effect of the core on the orbit parallel to the field, $V_{1}$, 
is relatively weak~\cite{Delos,Dando95,Dando96}. 
The periodic orbit labelling terminology of Ref.~\cite{H88} is used
throughout.

For the case of $s$-wave scattering,
each diffractive contribution in Eq.~(\ref{diffrac}) is:
\begin{equation}
dG = 
\sqrt{\hbar} (e^{2i\delta_{0}}-1) 
\left | \frac{2\pi}{m_{12}} 
\sin\frac{\theta_{i}}{2} \sin \frac{\theta_{f}}{2}\right |^{\frac12}
e^{i\left (S/\hbar - \mu\pi/2 -\pi/4 \right ) }
\label{diff2}
\end{equation}
and, in effect, represents the contribution of a pure diffractive orbit. 
Note the additional phase of $-\pi/4$ relative to an equivalent
geometric primitive periodic orbit.
 
Now we see that the amplitude of each non-hydrogenic primitive 
periodic orbit actually arises from the
interference between two contributions with the same action but 
different phase: a geometric one of the Gutzwiller form
weighted by the trace of the stability matrix, $M$, in the usual manner, 
{\em i.e.\/} for the $p$\/th orbit 
$A^p_H=(\pi S_p)/\sqrt{|2 -{\rm Tr} M_p|}$,  and 
a diffractive one, given by Eq.~(\ref{diff2}), following
the same path and of similar action but weighted by $1/\sqrt{m_{12}}$,
where $m_{12}$ is an element of $M$.  This  contrasts 
with the cardioid billiard where a typical contribution is 
either pure geometric or pure diffractive.

We can easily show that the fractional reduction of amplitude for a primitive 
periodic orbit of a non-hydrogenic atom relative to that of hydrogen is
\begin{equation}
A_{\delta}/A_H= \sqrt{1+4{\cal R}^2 \sin^2(\delta) \hbar  
-4{\cal R} \sin(\delta)\sin(\delta+\phi)\sqrt{\hbar}},
\label{frac}
\end{equation}
where, in general,
\begin{equation}
{\cal R}= A_H^{-1} \sqrt{\frac{32 \pi}{m_{12}}  
\sin(\theta_i/2) \sin(\theta_f/2)}.
\end{equation}
All parameters, {\em e.g.\/} initial and final angles, $\theta_i$, $\theta_f$,
refer to the particular primitive orbit under consideration.
The fractional reduction or increase is of $O(\sqrt{\hbar})$.
In general, $\phi=-\pi/4$. However, $R_1$ runs along
a boundary of the fundamental symmetry domain and so requires special 
treatment:  when stable, with winding number $\nu$, we find 
$\phi=-\pi/4-\nu\pi/2$ and 
${\cal R}=2 \sqrt{(\pi/m_{12})}\sin(\pi \nu)$.
The most important correction in Eq.~(\ref{frac}) is
the $\sqrt{\hbar}$ term. This is zero for $\delta=-\phi$ and positive
for $\sin(\delta +\phi) < 0$, leading to an {\em increased\/} amplitude for 
a primitive periodic orbit. In contrast,
in closed orbit theory, the main photoabsorption source term and the
core scattered terms do not have this $\pi/4$ de-phasing.

Harmonics of primitive orbits also have further 
contributions from product terms of $O(\hbar)$. 
However, product terms also give additional weak contributions at actions 
that are sums of periodic orbits. For example a two-orbit combination has 
amplitude
\begin{equation}
8 \pi^{2} (S_1+S_2) \frac{(e^{2i\delta} -1)^2}
{|m_{12}^{1}m_{12}^{2}|^{1/2}}
\left | \sin \frac{\theta_{i}^{1}}{2} \sin \frac{\theta_{f}^{1}}{2}
\sin\frac{\theta_{i}^{2}}{2} \sin \frac{\theta_{f}^{2}}{2}
\right | ^{1/2}.
\end{equation}
\begin{figure}[htpb]
\centerline{
\psfig{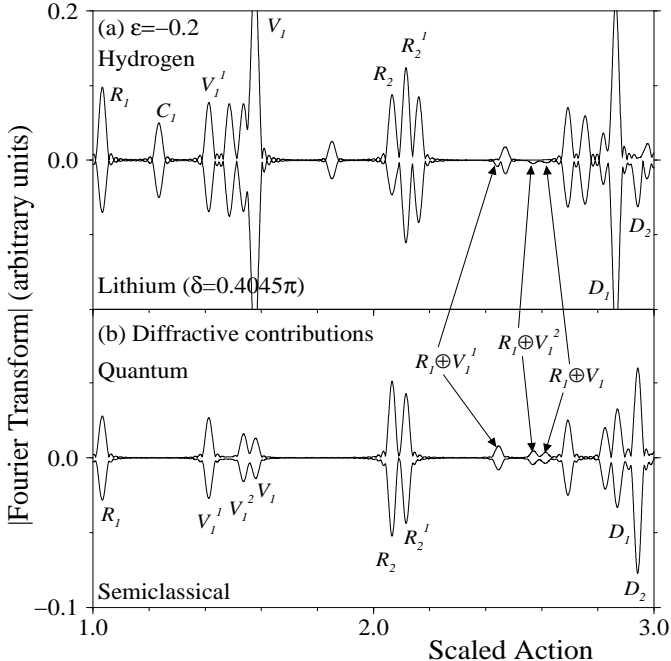}
}
\caption {(a) Comparison of Fourier transforms of the density of states 
for hydrogen and lithium ($\delta=0.4045\pi$) in a static magnetic field
at constant scaled energy $\varepsilon=-0.2$
from a fully quantal calculation with average $\hbar=1/90$.
Note the changes in amplitudes
of periodic orbits and new modulations due to diffractive orbits
in the lithium case.
(b) Comparison between quantal and semiclassical difference
spectra obtained by coherently subtracting the Fourier transforms
shown in (a).
This exposes the diffractive 
contributions to the spectrum and eliminates contributions from orbits which 
do not pass through the core. Shown are changes in periodic orbit
amplitudes due to diffraction, diffractive combinations of 
two periodic orbits and 
pure diffractive orbits, marked $D_1$ and $D_2$. 
Away from
bifurcations, which affect $V_1$ and $D_2$, the agreement between quantum
and semiclassical calculations is excellent.}
\label{Fig1}
\end{figure}

In Fig.~\ref{Fig1}(a) we show Fourier transforms of the oscillatory 
part of the even-$l$, $m=0$, eigenvalue spectra
for hydrogen and lithium ($\delta=0.4045\pi$) in a static 
magnetic field at constant scaled energy
$\epsilon=-0.2$ and with $n= \gamma^{-1/3}=\hbar^{-1}$ ranging from $60$ to
$120$. In Fig.~\ref{Fig1}(b)
we plot the `difference' spectrum obtained by coherently subtracting the 
Fourier transform of the hydrogenic spectrum from that of lithium;  
this exposes the diffractive contributions and eliminates contributions 
from periodic orbits which do not pass through the core.  
For comparison, we also plot a semiclassical
difference spectrum obtained by summing all terms of order $\sqrt{\hbar}$ 
and $\hbar$; agreement is excellent. The discrepancy in $D_2$
is due to the effects of bifurcations that are not taken into
account in the semiclassical calculation presented here.

We can see that for lithium
the amplitudes of $R_1$ and its harmonic $R_2$,
as well as $V_1^1$ and other orbits are substantially reduced.
There are additional small peaks which correspond
accurately to sums of periodic orbits. Importantly, there are
strong peaks (marked $D_1$ and $D_2$) which do not match any combination
of orbits. At these scaled actions ($S\simeq 2.87$ and $S\simeq 2.94$) 
we find  orbits
that are closed but not periodic. 
For hydrogen, only orbits that are
periodic in the fundamental symmetry domain contribute. Here we see
that pure diffractive orbits, such as $D_2$, can contribute to the
non-hydrogenic spectrum at
$O(\sqrt{\hbar})$ so are substantially stronger than combinations of 
orbits.  The peak at $S\simeq  2.87$ is due to an isolated closed orbit 
and is obtained almost exactly from Eq.~(\ref{diff2}) as seen in 
Fig.~\ref{Fig1}(b) (note that in Fig.~\ref{Fig1}(a) the peak associated 
with this  orbit is masked by the peak of a periodic orbit
which does not approach the nucleus). The peak at $S\simeq 2.94$
consists of contributions from a pair of non-isolated orbits close to a 
bifurcation so their contribution is over-estimated semiclassically. 
On examination of the diffractive orbits we find that they
correspond to the first closure of asymmetric periodic orbits,
some of which correspond to the $X_n$ series of `exotic orbits'~\cite{H88}.
In hydrogenic eigenvalue spectra such orbits only contribute at their
{\em full period,\/} whereas in the diffractive case they appear
at {\em closure.\/}

\begin{figure}[htbp]
\centerline{
\psfig{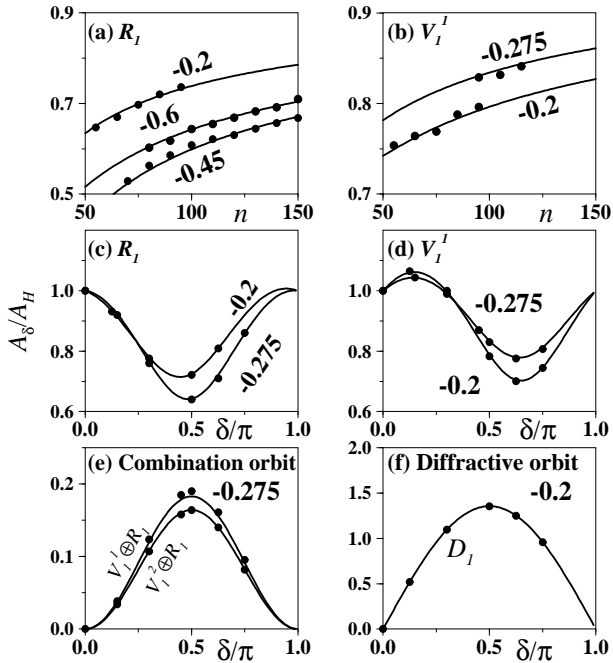}
}
\caption{Dependence of diffractive
contributions on $\hbar$ and $\delta$: comparison between
quantum results (full circles) and semiclassical formula,
Eq.~(\protect{\ref{frac}}), (solid line)
showing near exact agreement. The vertical axis represents the 
ratio of non-hydrogenic to hydrogenic amplitudes.  
(a)~Dependence of $A_{\delta}/A_H$ on $\hbar$ for $R_1$ at
$\epsilon=-0.2$, $-0.6$ and $-0.45$.
(b) Same as (a) for $V_1^1$ at $\epsilon=-0.2$, and
$-0.275$. 
(c)~Dependence of $A_{\delta}/A_H$ on quantum defect, $\delta$, for
$R_1$.
(d)~Same as (c) for $V_1^1$:  note the
de-phasing relative to $R_1$ and that the amplitude exceeds the hydrogenic
value for $\delta < \pi/4$. (e)~and~(f) dependence of contributions of 
combination and diffractive orbits on $\delta$. Here the ratio 
relative to the first contribution of the circular orbit, $C$, is shown. 
As predicted by the theory, combination orbits show a 
$\sin^2 \delta$ behaviour while diffractive orbits follow 
a $\sin \delta$ curve.}
\label{Fig2}
\end{figure}

We have carried out a detailed study of these effects for several  
scaled energies to study the $\hbar$ and $\delta$ dependence of
the diffractive effects. 
In Figs~\ref{Fig2}(a)--(d) we compare the fractional change relative
to hydrogen between the fully quantal and
semiclassical expressions for $R_1$ and $V_1^1$.
The agreement is very good. For the $\hbar$ dependence there are 
fewer points for $V^1_1$ since a wide spectral range is required to 
resolve it from a nearby orbit. An especially interesting feature is the
de-phasing of $R_1$ relative to $V^1_1$ seen in Figs~\ref{Fig2}(c) and 
\ref{Fig2}(d).
The diffractive contribution to $V^1_1$ is $-\pi/4$ out of phase
with the geometric term. As a result the amplitude exceeds
that of hydrogen for $\delta \lesssim \pi/4$ and is minimal at 
$\delta \simeq 0.65\pi$.
In contrast, the geometric and diffractive contributions for $R_1$ are
almost in phase at $\epsilon =-0.275$ and remain so for a wide range 
of scaled energies about $\epsilon \simeq -0.3$, where the orbit
undergoes its $2:1$ resonance with $\nu\simeq 0.5$.

In Figs~\ref{Fig2}(e) and ~\ref{Fig2}(f) we investigate the combination
orbits and the diffractive orbit that appears at $\epsilon =-0.2$
for $S\simeq 2.87$. In this case we plot the ratio of amplitudes
relative to the first peak of $C$, the circular orbit, a periodic
orbit that does not pass through the nucleus and hence is unaffected 
by the diffraction.  In both cases the agreement is very good. 

In conclusion, we have shown that periodic orbit theory (the GTF) 
may be applied to all singly excited atoms as successfully as for hydrogen by 
bringing in the effects of QDT in the form of diffractive corrections.
Also, to our knowledge, this is the first demonstration of a diffractive
effect in a real system, since previously diffraction has only been applied 
to model problems such as billiards. Although we show explicit results for 
$s$-wave scattering in lithium and helium, our method is applicable
generally to other Rydberg atoms and molecules in external fields.

An interesting recent calculation~\cite{Main} treated the photoabsorption
of general atoms  within the framework of the standard
theory  using a model potential. Then, the observed closed orbit modulations
were modelled by superposing thousands of very unstable orbits.
Hence, the issue of whether the dynamics of non-hydrogenic atoms at
moderate scaled energies is an instance of chaos ({\em i.e.\/} very unstable 
motion) or an effect `beyond periodic orbits', such as 
diffraction, remains open. Our work addressed this issue.

Currently, there is added interest in diffractive systems
since they have very recently been associated with a new class
of intermediate level statistics (`Half-Poisson')~\cite{Bog97}. 
Recently, eigenvalue statistics for rubidium in fields were investigated
experimentally~\cite{Held} and shown to be nearer the GOE limit
than comparable hydrogenic results. Hence, given that spectroscopic
resolution exceeds mean level spacing, experimental verification of 
diffractive effects in the eigenvalue spectrum, for example the presence 
of the `D' modulations, is in principle possible.

We are indebted to E.B.\ Bogomolny, D.\ Delande and J.B.\ Delos
for helpful advice and discussions.
The authors acknowledge funding from the EPSRC.
SMO is supported by an EPSRC studentship.

\end{document}